\documentclass[pra,aps,reprint,superscriptaddress,longbibliography,floatfix]{revtex4-1}
\usepackage{times}
\usepackage{epsfig}
\usepackage{amsfonts}
\usepackage{amsmath}
\usepackage{amssymb}
\usepackage{amsthm}
\usepackage{color}
\usepackage{multirow}
\usepackage[normalem]{ulem}
\newcommand{\stkout}[1]{\ifmmode\text{\sout{\ensuremath{#1}}}\else\sout{#1}\fi}
\usepackage{latexsym}
\usepackage{mathrsfs}
\usepackage{verbatim}
\usepackage{mathpazo}
\usepackage[T1]{fontenc}
\usepackage[colorlinks=true,linkcolor=blue,citecolor=magenta,urlcolor=blue]{hyperref}
\usepackage{apptools}
\usepackage{chngcntr}

\AtAppendix{\counterwithin{lemma}{section}}

\newcommand{\be}{\begin{equation}}
\newcommand{\ee}{\end{equation}}
\newcommand{\beq}{\begin{eqnarray}}
\newcommand{\eeq}{\end{eqnarray}}

\DeclareMathOperator{\tr}{Tr}

\def\squareforqed{\hbox{\rlap{$\sqcap$}$\sqcup$}}
\def\qed{\ifmmode\squareforqed\else{\unskip\nobreak\hfil
\penalty50\hskip1em\null\nobreak\hfil\squareforqed
\parfillskip=0pt\finalhyphendemerits=0\endgraf}\fi}
\def\endenv{\ifmmode\;\else{\unskip\nobreak\hfil
\penalty50\hskip1em\null\nobreak\hfil\;
\parfillskip=0pt\finalhyphendemerits=0\endgraf}\fi}

\newcommand{\blk}{\color{black}}

\newtheorem{lemma}{Lemma}
\newtheorem{proposition}{Proposition}
\newtheorem{fact}{Corollary}
\newtheorem{result}{Result}

\begin{document}

\title{Preparation contextuality as an essential feature underlying quantum communication advantage}

\author{Debashis Saha}
\affiliation{Institute of Theoretical Physics and Astrophysics, National Quantum Information Centre, Faculty of Mathematics, Physics and Informatics, University of Gda\'{n}sk, 80-952 Gda\'{n}sk, Poland}
\affiliation{Center for Theoretical Physics, Polish Academy of Sciences, Al. Lotnik\'{o}w 32/46, 02-668 Warsaw, Poland}

\author{Anubhav Chaturvedi}
\affiliation{Institute of Theoretical Physics and Astrophysics, National Quantum Information Centre, Faculty of Mathematics, Physics and Informatics, University of Gda\'{n}sk, 80-952 Gda\'{n}sk, Poland}

\begin{abstract}
The study of ontology (hidden variables) provides for a vital ground on which significant non-classical features of quantum theory are revealed. One such non-classical ontic-feature is preparation contextuality ($\mathcal{PC}$) and advantage in oblivious communication tasks is one of its operational signatures.  This article primarily pursues the ontic-feature underlying quantum advantage in communication complexity ($\mathcal{CC}$). We construct oblivious communication tasks tailored to given $\mathcal{CC}$ problems. We upper-bound the classical success probability of these oblivious communication tasks, obtaining preparation non-contextual inequalities. 
We use the very states and measurements responsible for advantage in $\mathcal{CC}$ problems along with the orthogonal mixtures of these states to orchestrate an advantageous protocol for the associated oblivious communication tasks and the violation of the associated inequalities, thereby unveiling $\mathcal{PC}$. 
To showcase the vitality of our results, we find a criterion for unbounded violation of these inequalities and demonstrate the same for two widely studied $\mathcal{CC}$ problems. 
\end{abstract}

\maketitle
\section{Introduction} 
Quantum resources coupled with ingenious quantum protocols have outshone their classical counterparts in a wide range of computation, communication, and information processing tasks. But there is little insight into what makes quantum theory stand out. The theory-specific features such as superposition do not make insightful answers for they are cyclic in the sense that they refer back to the operational quantum formalism which was a priori responsible for the advantageous predictions. 
Therefore, any comprehensive approach to this question must arguably invoke a ground common to both classical and quantum theories, on which non-classical features of the latter are unveiled.
The study of hidden variables (ontology) provides for such a ground. Any ontological model that seeks to explain the predictions of operational quantum formalism must have certain non-classical features \cite{book,cabello}. 
Introduced in \cite{Spekkens2005}, the ontic-feature of preparation contextuality ($\mathcal{PC}$) discards any preparation non-contextual ($\mathcal{PNC}$) models as viable ontological descriptions of an operational theory. An ontological model is said to be $\mathcal{PNC}$ if it assigns identical ontic-distributions to operationally indistinguishable preparations \cite{Spekkens2005}. Quantum theory manifests preparation contextuality ($\mathcal{PC}$), i.e., it postulates certain operationally indistinguishable preparations which must have non-identical underlying ontic-distributions. Quantum protocols siphon this ontological distinguishability to an advantage in oblivious communication ($\mathcal{OC}$) tasks i.e. \textit{any} advantage in $\mathcal{OC}$ tasks witnesses $\mathcal{PC}$ \cite{Spekkens09,shp,chutiyapa,comment}. \\
One of the predominant manifestations of quantum communication advantage is captured in communication complexity ($\mathcal{CC}$).
The notion of $\mathcal{CC}$, introduced in the seminal paper \cite{YaoCom}, is an important aspect of complexity theory, which quantifies the amount of communication required for distributed computation. Apart from mainstream applications in algorithmic mechanism design, game theory and cryptography, lower bounds in $\mathcal{CC}$ can be used to prove lower bounds in decision tree complexity, data structures, space-time trade-offs for Turing machines and more \cite{ccbook}. Quantum resources and strategies have demonstrated supremacy in a multitude of $\mathcal{CC}$ problems \cite{brassard,wolf,raz,review,BCW,BCWW}. 
In this article, we substantiate a fundamental link between quantum $\mathcal{CC}$ advantage and $\mathcal{PC}$. Specifically, we establish that quantum advantage in $\mathcal{CC}$ manifests $\mathcal{PC}$. We begin by constructing an $\mathcal{OC}$ task tailored to a given instance of the generic $\mathcal{CC}$ problem. We orchestrate advantageous quantum strategies for the constructed $\mathcal{OC}$ tasks based on advantageous $(i)$ one-way prepare and measure quantum $\mathcal{CC}$ protocols, $(ii)$ two-way multi-round quantum $\mathcal{CC}$ protocols and, $(iii)$ entanglement assisted classical communication $\mathcal{CC}$ protocols. These $\mathcal{OC}$ strategies utilize the same quantum set-up responsible for advantage in the $\mathcal{CC}$ task. Specifically, we provide a family of $\mathcal{PNC}$ inequalities tailored to $\mathcal{CC}$ tasks and show that quantum $\mathcal{CC}$ advantage implies a violation of these inequalities, subject to certain conditions. Additionally, we obtain a criterion for unbounded violation of these $\mathcal{PNC}$ inequalities and demonstrate it for two widely studied $\mathcal{CC}$ problems with exponential quantum advantage. We present an alternative construction of the $\mathcal{OC}$ task and discuss the potential extension of our results to general probabilistic theories. Next, we use the machinery thus developed to provide a complete proof of the fact (originally stated in \cite{chutiyapa}) that violation of (spatial or temporal) Bell inequalities \cite{Bell66,bellreview,LP} implies an advantage in an associated $\mathcal{OC}$ task.
Finally, we conclude with a discussion of the implications of this work.

\section{Primitives}
In this section, we lay down the framework we employ in our investigation. Specifically, we introduce the generic formulations of $\mathcal{CC}$ problems and $\mathcal{OC}$ tasks, which form the key subjects of this article.
\subsection{Communication complexity problem}
We begin with briefly introducing the generic formulation of $\mathcal{CC}$ problem.
A typical $\mathcal{CC}$ problem entails two parties Alice and Bob, with inputs $x\in [n_x]$, $y\in [n_y]$ (where $[n]=\{0,1,\ldots,n-1\}$) respectively, distributed according to a prior probability distribution $p(x,y)$. Their task is to compute the value of a binary output bivariate function, $f(x,y) \colon [n_x]\times[n_y] \to \{0,1\}$ by exchanging messages. Without loss of generality, we assume that Bob guesses the value of $f(x,y)$ and his guess is stored in an output bit $z\in\{0,1\}$. They achieve success with probability, 
\be
p =  \sum_{x,y} p(x,y) p(z=f(x,y)|x,y).\ee
There are two inter-convertible metrics to gauge their performance: $(i)$ maximal achievable success probability ($p_{\mathcal{C}_d}$ for classical resources and $p_{\mathcal{Q}_d}$ for quantum resources) given a bounded amount of communication (bounded dimension $d$ of the communicated system), and $(ii)$ amount of communication (usually quantified in bits, denoted by $\mathcal{C}(f,p_{\mathcal{S}})$ or qubits, denoted by $\mathcal{Q}(f,p_{\mathcal{S}})$) required to achieve to achieve a specified probability of success (denoted by $p_{\mathcal{S}}$). Quantum $\mathcal{CC}$ advantage implies $p_{\mathcal{Q}_d}>p_{\mathcal{C}_d}$ or alternatively $\mathcal{Q}(f,p_{\mathcal{S}})<\mathcal{C}(f,p_{\mathcal{S}})$.

\subsection{Oblivious communication task}\label{sec:oc}
For this article, we need only invoke a subclass of general $\mathcal{OC}$ tasks (introduced in \cite{shp}) wherein Alice's (sender) input comprises of a pair $a=(a_1,a_2)$ with $a_1\in [n_{a_1}]$, $a_2\in [n_{a_2}]$. Bob (receiver) gets an input $b \in [n_b]$ and yields an output $c\in [n_c]$. The inputs are distributed according to a prior probability distribution $p(a,b)$ with an additional condition $p(a_2|a_1,b)=p(a_2|a_1)$. Their task is to guess the value of a function $g(a,b):[n_a] \times [n_b] \to [n_c]$. In contrast to $\mathcal{CC}$ problems, there is no restriction on the amount of communication.  The communication is constrained to be completely oblivious to the value of $a_1$. They achieve success with probability defined as $p = \sum_{a,b} p(a,b) p(c=g(a,b)|a,b)$.\\
In a classical $\mathcal{OC}$ protocol Alice prepares the message $m$ employing an encoding scheme $\mathcal{E}$ which comprises of conditional probability distributions of the form $p_{\mathcal{E}}(m|a)$. Bob outputs $c$ based on his input $b$ and the message $m$ using a decoding scheme $\mathcal{D}$ entailing conditional probability distributions $p_{\mathcal{D}}(c|b,m)$. The oblivious constraint for classical encoding schemes $\mathcal{E}$ reads as, 
\be
\forall m, \forall a_1,{a_1}'\in[n_{a_1}], \  p_{\mathcal{E}}(m) =p_{\mathcal{E}}(m|a_1)= p_{\mathcal{E}}(m|{a_1}'),
\ee where $p_{\mathcal{E}}(m|a_1)=\sum_{a_2} p(a_2|a_1) p_{\mathcal{E}}(m|a_1,a_2)$.
This condition ensures that the same classical mixture is prepared for all values of $a_1$. The expression for maximal classical success probability is,
\begin{align} \label{NCB0} \nonumber
p_{\mathcal{NC}} = & \max_{\{\mathcal{E}\}\{\mathcal{D}\}} \bigg\{ \sum_{m} \sum_{b} p(b) \\ & \bigg(\sum_{a} p(a|b) p_{\mathcal{E}}(m|a) p_{\mathcal{D}}(g(a,b)|b,m) \bigg)\bigg\},
\end{align} 
where the message $m$ can take arbitrary number of distinct values. We use  the subscript $\mathcal{NC}$ to reflect the fact that for $\mathcal{OC}$ tasks the maximal classical success probability is the same as the maximal $\mathcal{PNC}$ success probability \cite{chutiyapa,shp}.

On the other hand, quantum strategy for a $\mathcal{OC}$ task involves Alice transmitting states of arbitrary dimension, $\rho_a$ for input $a$, such that the same mixed state $\rho$ is prepared for all values of $a_1$ i.e., $\forall a_1,  \ \sum_{a_2}p(a_2|a_1)\rho_{a_1,a_2}=\rho$. This ensures adherence to the oblivious condition. Upon receiving input $b$, Bob performs measurement $\{M^b_c\}$ (where $\sum_c M^b_c=\mathbb{I}$) on the transmitted system. The average success probability is given by the expression $p_{\mathcal{Q}}=\sum_{a,b} p(a,b) \tr(\rho_a M^b_{c=g(a,b)})$. \\

\section{Advantage in $\mathcal{CC}$ implies advantage in $\mathcal{OC}$}
In this section we present our main results. First, we make a couple of essential observations concerning the maximal classical success probability of $\mathcal{OC}$ tasks.
Next, we construct an $\mathcal{OC}$ task tailored to a given instance of generic $\mathcal{CC}$ problem described in the previous section. We then formulate a $\mathcal{PNC}$ inequality by obtaining an upper-bound on the classical success probability of the $\mathcal{OC}$ task. We utilize the very resources responsible for quantum advantage in the given $\mathcal{CC}$ problem (pertaining to without prior entanglement $i.$ one-way prepare and measure protocols, $ii.$ two-way multi-round protocols and, $iii.$ entanglement assisted classical communication protocols) to orchestrate an advantageous quantum protocol for the associated $\mathcal{OC}$ task, thereby demonstrating the violation of the $\mathcal{PNC}$ inequality. Further, we present two instances of unbounded violations of $\mathcal{PNC}$ inequalities based on $\mathcal{CC}$ problems with exponential quantum advantage.  
Finally, we provide an alternative construction of $\mathcal{OC}$ task tailored to given $\mathcal{CC}$ problems and discuss the persistence of our results in general probabilistic theories. 

\subsection{Bounding classical success in $\mathcal{OC}$ tasks}

In general, finding maximal classical success probability for $\mathcal{OC}$ tasks is an arduous task as $(i)$ the dimension of the message is unbounded and, $(ii)$ the encoding scheme may be probabilistic. In lieu of these issues we employ the following lemmas (based on the observation in \cite{shp}) to facilitate an upper bound on classical success probability of the $\mathcal{OC}$ task,

\begin{lemma} \label{le:UpperBound}
For an instance of the subclass of $\mathcal{OC}$ tasks defined in Section \ref{sec:oc}, the classical success probability $p_\mathcal{NC}$ is upper bounded in the following way,
\begin{align} \label{ncb1}
p_{\mathcal{NC}} & \leq \max_{\{q_{a_1,a_2}\}} \bigg\{ \sum_b p(b) \max_c \bigg\{ 
\sum_{a_1,a_2} p(a_1|b) q_{a_1,a_2} \delta_{c,g(a,b)} \bigg\} \bigg\}
\end{align} 
where the outer maximization is over a set of variables $\{q_{a_1,a_2}\}$  satisfying the conditions,
\be \label{q1}
 q_{a_1,a_2} \geq0, \ \sum_{a_2} q_{a_1,a_2} = 1. \ee 
\end{lemma}
\begin{lemma} \label{le:extremal} The set of valid assignments of $\{q_{a_1,a_2}\}$ satisfying the linear constraints \eqref{q1} form a convex polytope. The extremal points of this polytope resemble deterministic probability distributions, i.e., any extremal point $\{q^{ext}_{a_1,a_2}\}$ is of the form: for each $a_1$, $q^{ext}_{a_1,a_2}=0$ for all values of $a_2$ except a specific $\tilde{a_2}$ for which $q^{ext}_{a_1,\tilde{a_2}}=1$. \blk
\end{lemma}
The proofs have been deferred to the Appendix. It follows from \textit{Lemma} \ref{le:extremal} that the outer maximization in \eqref{ncb1} can be performed by evaluating the contained expression at each extremal point of convex polytope formed by the valid assignments of $\{q_{a_1,a_2}\}$. Let the extremal point yielding the maximal value be $\{q^{ext,max}_{a_1,a_2}\}$. This extremal point without loss of generality entails for each $a_1$, an $\tilde{a_2}$ where $q^{ext,max}_{a_1,\tilde{a_2}}=1$. Let for each $a_1$, $\tilde{a_2}=e_{a_1}$, then we have $q^{ext,max}_{a_1,a_2}=\delta_{a_2,e_{a_1}}$. Similarly, for the inner maximization suppose that for this extremal point, for each $b$ the maximal value of $ 
\sum_{a_1,a_2} p(a_1|b) q_{a_1,a_2} \delta_{c,g(a,b)}$ is obtained for $c=c_b$. Consequently, we arrive at the following distilled re-expression of \eqref{ncb1},
\be \label{optnc*}
p_{\mathcal{NC}} \leq \sum_{b} p(b) \sum_{a_1,a_2} p(a_1|b)\ \delta_{e_{a_1},a_2}\ \delta_{c_b,g(a,b)}.
\ee

\subsection{Tailoring $\mathcal{OC}$ tasks to given $\mathcal{CC}$ problems and $\mathcal{PNC}$ inequality}
\begin{figure}[http]
\centering
\includegraphics[width=0.73\columnwidth]{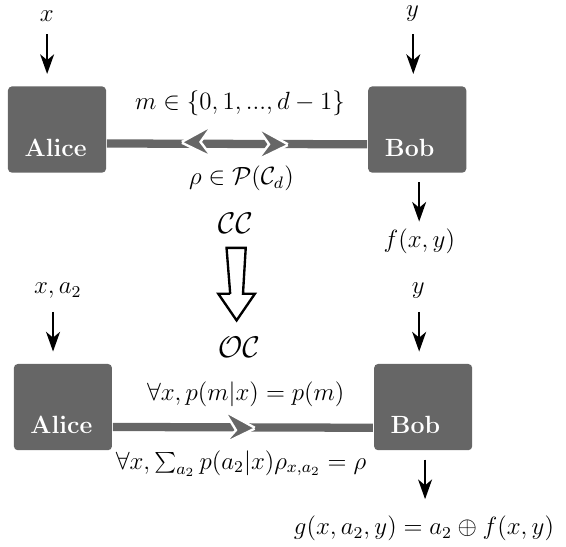}
\caption{Construction of $\mathcal{OC}$ task based on a given $\mathcal{CC}$ task. Notice while the amount of communication in the $\mathcal{CC}$ task by the dimension $d$ of the physical system, there is no such constraint on communication in the $\mathcal{OC}$. Instead the communication is restricted so as not to reveal any information about the oblivious variable $x$.}
\label{fig1}
\end{figure}

We now present the key ingredient of our modus-operandi, an $\mathcal{OC}$ task tailored to a given $\mathcal{CC}$ problem.
Given an instance of the generic $\mathcal{CC}$ problem described above, we construct the following $\mathcal{OC}$ task (see Fig.\ref{fig1}),
\beq \label{oc}
&a = (a_1=x,a_2), \ b = y,\ c=z,\nonumber \\
&p(a,b) = p(x,a_2,y) = p(y)p(x|y)p(a_2|x), \nonumber \\
&\text{where } a_2 \in \{0,1\}, \ p(a_2|x) = 
 \begin{cases} 
    \frac{1}{d}, &\text{if } a_2=0\\
    \frac{d-1}{d},&\text{if } a_2=1
  \end{cases} \nonumber \\
& \text{and } g(x,a_2,y) = f(x,y) \oplus a_2.
\eeq
Recall, that in the $\mathcal{OC}$ task the oblivious condition constrains the communicated system to not carry any retrievable information about $x$. 

Next, by the means of the following proposition which upper-bounds the classical success probability for the constructed $\mathcal{OC}$ task, we present a family of $\mathcal{PNC}$ inequalities tailored to $\mathcal{CC}$ problems.
\begin{proposition} \label{prop:pncIneqMain} The $\mathcal{PNC}$ success probability of the $\mathcal{OC}$ task described in \eqref{oc} is upper bounded by the maximal classical success probability of the $\mathcal{CC}$ problem wherein Alice is restricted to communicate a two-leveled system, i.e. 
\be \label{PNCineq1}
p_\mathcal{NC} \leq p_{\mathcal{C}_2}.
\ee
\end{proposition}
\begin{proof} The proof involves obtaining an upper-bound for the classical success probability of the $\mathcal{OC}$-task (constructed above) with the help of \textit{Lemma} \ref{le:UpperBound} and \textit{Lemma} \ref{le:extremal}. We then show that this upper-bound forms a viable (not necessarily optimal) classical success probability for the original $\mathcal{CC}$ problem whilst the dimension of the message is restricted to two. Note that when the communication is restricted to be at-most two-dimensional, two-way multi-round $\mathcal{CC}$ protocols are equivalent to one-way $\mathcal{CC}$ protocols wherein only Alice is allowed to communicate a two-level message $m\in \{0,1\}$ to Bob, deeming this inequality to be independent of the choice of protocol. \\ 
The expression for maximal classical success probability of the $\mathcal{CC}$ task when Alice is restricted to transmit a bit of communication $p_{\mathcal{C}_2}$ reads, 
\beq \label{PCOWCC1} \nonumber
&p_{\mathcal{C}_2}  = &\max_{\{{E}\}\{{D}\}} \bigg\{  \sum_{y} p(y) \bigg( \\
& &\sum^{1}_{m=0} \sum_{x} p(x|y) p_{E}(m|x) p_{D}(z=f(x,y)|y,m) \bigg)\bigg\},
\eeq
where Alice's encoding scheme $E$ entails conditional probability distributions of the form $p_{E}(m|x)$ and Bob's decoding scheme $D$ entails conditional probability distributions of the form $p_{D}(z|y,m)$.
On the other hand, it follows from \eqref{optnc*} that the classical success probability of the $\mathcal{OC}$ task is upper bounded as follows,
\beq \label{desiredThesis}
&p_{\mathcal{NC}} \leq & \sum_y p(y) \sum_{x,a_2} p(x|y) \delta_{e_x,a_2} \ \delta_{c_y,a_2\oplus f(x,y)} \nonumber \\
&=&  \sum_y p(y) \sum_{x,a_2} p(x|y) \delta_{e_x,a_2} \ \delta_{a_2,c_y \oplus f(x,y)} \nonumber \\
&=& \sum_{x,y} p(x,y) \delta_{e_x,c_y \oplus f(x,y)}.
\eeq
To complete the proof we demonstrate that this upper bound (RHS of \eqref{desiredThesis}) is achievable in the $\mathcal{CC}$ task employing a two-leveled message $m\in \{0,1\}$. To this end, we present the following classical $\mathcal{CC}$ protocol,
\begin{align}
p_{E}(m|x) = \delta_{m,e_x} \ , \ p_{D}(z|y,m) = \delta_{z, c_y \oplus m}. 
\end{align}
Inserting this strategy in \eqref{PCOWCC1}, one obtains
\beq 
&p_{\mathcal{C}_2}\ &\geq \sum_y p(y) \sum_{x,m} p(x|y) \delta_{m,e_x} \ \delta_{f(x,y),c_y\oplus m} \nonumber \\
&& = \sum_{x,y} p(x,y) \delta_{e_x,c_y \oplus f(x,y)}\ , 
\eeq
which together with \eqref{desiredThesis} yields the desired thesis \eqref{PCOWCC1}. 
\end{proof}
\subsection{Violation of $\mathcal{PNC}$ inequality from advantageous quantum $\mathcal{CC}$ protocols}
Notice that up-until this point our results are independent of the specifics of the $\mathcal{CC}$ protocol including the restriction on the amount of communication, but depend only on the problem itself.
Now we take three distinct classes of the advantageous quantum $\mathcal{CC}$ protocols and based on these, we construct quantum strategies for the $\mathcal{OC}$ task to demonstrate the violation of the associated $\mathcal{PNC}$ inequalities.
\subsubsection{One-way prepare and measure quantum $\mathcal{CC}$ protocols}
One-way quantum $\mathcal{CC}$ protocols without prior entanglement  are commonly referred to as \textit{prepare and measure protocols}. In such protocols, Alice's state (a qu$d$it $\rho_x$ for input $x$) preparation and transmission is followed by a binary outcome measurement ($\{M^y_z\}$ upon receiving input $y$) at Bob's end. The quantum success probability is expressed as,
\be \label{pQd1}
p_{\mathcal{Q}_d} = \sum_{x,y} p(x,y) \tr(\rho_x M^y_{z=f(x,y)}).
\ee 
Notice that here, quantum success probability $p_{\mathcal{Q}_d}$ is not required to be maximal. Now we present our result concerning $\mathcal{PC}$ manifest in advantageous prepare and measure quantum $\mathcal{CC}$ protocols.
\begin{result} \label{result:1}
Given a prepare and measure quantum $\mathcal{CC}$ protocol, an advantage is obtained in the $\mathcal{OC}$ task described in \eqref{oc} ($p_\mathcal{Q}>  p_\mathcal{NC}$)  whenever the following condition holds,
\be \label{c7}
\frac{1}{d}(2p_{\mathcal{Q}_d} + d-1 - \chi) > p_{\mathcal{C}_2} ,
\ee 
where $\chi = \sum_{x,y} p(x,y)\tr\left( M^y_{z=f(x,y)}\right)$ and $\{M^y_{z}\}$ are Bob's measurements employed in quantum $\mathcal{CC}$ protocol.
\end{result}
\begin{proof} Our quantum strategy for the $\mathcal{OC}$ task described in \eqref{oc}, involves Alice preparing the same states (as in the quantum $\mathcal{CC}$ protocol described above) when $a_2=0$ i.e. $\rho_{x,a_2=0}=\rho_{x}$ and their orthogonal mixtures when $a_2=1$ i.e. $\rho_{x,a_2=1}=\frac{\mathbb{I}-\rho_{x}}{d-1}$. Alice's preparations are therefore oblivious to $x$, as $\forall x: \ \sum_{a_2} p(a_2|x) \rho_{x,a_2}=\frac{\mathbb{I}}{d}$. Bob's measurements remain unaltered from the quantum $\mathcal{CC}$ protocol.  Plugging the expressions of $p(x,a_2,y)$ from \eqref{oc} and $p_{\mathcal{Q}_d}$ from \eqref{pQd1}, we obtain the following success probability for this strategy, 
\begin{align} \label{myLove}
p_{\mathcal{Q}} & = \sum_{x,a_2=0,y} p(x,a_2,y) \tr(\rho_x M^y_{z=f(x,y)}) \nonumber \\ & \hspace{23pt} + \sum_{x,a_2=1,y} p(x,a_2,y) \tr\left(\frac{\mathbb{I} - \rho_x}{d-1}M^y_{z=1\oplus f(x,y)} \right) \nonumber \\  & = \frac{1}{d}\left( 2p_{\mathcal{Q}_d} + d-1 - \chi \right), 
\end{align}
where $\chi = \sum_{x,y} p(x,y)\tr\left( M^y_{z=f(x,y)}\right)$. Now our desired result simply follows from \eqref{PNCineq1}.
\end{proof}
 Now, given that the $\mathcal{CC}$ protocol under consideration is advantageous, i.e. $p_{\mathcal{Q}_d}>p_{\mathcal{C}_d}$, if follows that a quantum advantage in the $\mathcal{OC}$ task is obtained  ($p_{\mathcal{Q}} > p_{\mathcal{NC}}$) whenever the following holds,
 \be \label{mod1}
\frac{1}{d}(2p_{\mathcal{C}_d} + d-1 - \chi) \geq p_{\mathcal{C}_2} .
\ee 
To aid intuition and accessibility we simplify the above condition \eqref{c7} employing two lemmas (the proofs are deferred to the Appendix):
\begin{lemma} \label{le:3} For a given prepare and measure quantum $\mathcal{CC}$ protocol the following holds,
\be \label{chi}
\chi \leq d p_G,
\ee
where  $\chi = \sum_{x,y} p(x,y)\tr\left( M^y_{z=f(x,y)}\right)$, $d$ is dimension of the communicated system and $p_G$ is guessing probability without communication.
\end{lemma}

\begin{lemma} \label{le:4} Given a $\mathcal{CC}$ problem and a classical protocol using a two-leveled classical message with a success probability $p_{\mathcal{C}_2}$, the success probability of a protocol using a $d$-leveled classical message is lower bounded in the following way,
\be \label{pump}
p_{\mathcal{C}_d} \geq 1 - \exp{\left(-\frac{\log{d}}{2p_{\mathcal{C}_2}}(p_{\mathcal{C}_2}-\frac{1}{2})^2\right)}.
\ee 
\end{lemma}
\begin{fact}
By substituting the upper bound of $\chi$ from \eqref{chi} in the condition \eqref{c7}, we find that $p_{\mathcal{Q}}>p_{\mathcal{NC}}$ whenever $p_{\mathcal{Q}_2}>p_{\mathcal{C}_2}$ in any $\mathcal{CC}$ task with $p_G=\frac{1}{2}$. 
\end{fact}
\begin{fact}
By imposing {\it Lemma} \ref{le:3}-\ref{le:4} into \eqref{mod1}, we find that $p_{\mathcal{Q}}>p_{\mathcal{NC}}$ whenever the following condition holds,
\be 
\label{CC}
d(p_{\mathcal{C}_2} + p_G -1) + 2\exp{\left(-\frac{\log{d}}{2p_{\mathcal{C}_2}}(p_{\mathcal{C}_2}-\frac{1}{2})^2\right)} \leq 1.
\ee  
\end{fact} 
Notice, \eqref{CC} relies only on classical success probability of the $\mathcal{CC}$ task with a two-leveled message $p_{\mathcal{C}_2}$ and success probability of the $\mathcal{CC}$ task without any communication $p_G$. This in-turn deems \eqref{CC} to be independent of the specifics of the implementation of classical or quantum $\mathcal{CC}$ protocols including the dimension of the communicated system.   \\

\subsubsection{Two-way multi-round quantum $\mathcal{CC}$ protocols}
Even though one-way $\mathcal{CC}$ protocols form a predominant subclass of quantum $\mathcal{CC}$ protocols, two-way multi-round $\mathcal{CC}$ protocols employ relatively more involved features of quantum theory to facilitate  an advantage \cite{raz}. In two-way multi-round $\mathcal{CC}$ protocols, Alice and Bob have access to local quantum memories and exchange messages over multiple rounds of communication. In each round they use local operations to store an imprint of the message on their respective local memories and prepare a message for the next round. This results in complex pre-measurement states wherein Alice's local memory may be entangled with Bob's local memory. Remarkably, our results hold intact for quantum advantage in $\mathcal{CC}$ tasks obtained via two-way multi-round $\mathcal{CC}$ protocols. \\
We start by presenting a general two-way multi-round $\mathcal{CC}$ protocol denoted by $\mathcal{P}$ (first described in \cite{pnas}). Alice and Bob have access to some quantum memory, the states of respective quantum memory in the round $r$ are symbolized by $A^{x,y}_r$ and $B^{x,y}_r$. These symbols serve for the convenience of description and for mere subscripts of the quantum state $\rho$.
Each round consists of transmission of a message from Alice to Bob and back. We symbolize the communicated quantum system  from Alice to Bob and from Bob to Alice in the round $r$ by  $\alpha_r$ and $\beta_r$, respectively.
Let the total number of rounds be $R$.
The protocol proceeds as follows,  
\begin{enumerate}
\item Depending on the input $x$, Alice applies a local operation $U^x_1$ on the joint system of her initial memory $A_0$ and the blank message 
$\alpha$ to obtain an updated combined state $\rho_{\alpha_1,A^x_1}$ with local memory $A^{x}_1$ and the message $\alpha_1$. Alice then sends the message i.e. the reduced state $\rho_{\alpha_1}$  to Bob. In general the updated local memory and the message may now be entangled.
\item Depending on the input $y$, Bob applies a local operation $U^y_1$ on the joint system of his local memory $B_0$ and the message from Alice $\alpha_1$ to obtain his updated combined system $\rho_{\beta_1,B^{x,y}_1}$ with local memory $B^{x,y}_1$ and the message $\beta_1$ which is then communicated back to Alice.  As a result, Bob's local memory $B_1^{x,y}$ may be entangled with Alice's local memory $A^x_1$.
\item This marks the completion of the first round. Alice and Bob repeat these steps for $R-1$ rounds. In the last round $(r=R)$ upon receiving the message from Alice ($\alpha_R$) instead of sending a message back to Alice, Bob performs the measurement $\{M^y_z\}$ on the joint system of the message and Bob's local memory from the previous round ($B^{x,y}_{R-1}$).
\end{enumerate}
Given an upper bound on total dimension of communication $d$, they achieve success with probability 
$p_{\mathcal{Q}_d}=\sum_{x,y}p(x,y)\tr(\rho_{\alpha_R,B^{x,y}_{R-1}}M^y_{z=f(x,y)}),$ where $\rho_{\alpha_R,B^{x,y}_{R-1}}$ is the reduced density matrix corresponding to the joint system of the message from Alice ($\alpha_R$) and Bob's local memory from the penultimate round $B^{x,y}_{R-1}$.  \\
To what follows, it is crucial to obtain an upper-bound on the dimension of Bob's pre-measurement state. We achieve this by employing the following steps, 
\begin{enumerate}
\item Following the methodology in \cite{pnas}, we first convert a given two-way multi-round quantum communication protocol $\mathcal{P}$ utilizing $\log_2{d}$-qubit (i.e. $d$ dimensional communication) communication to another protocol $\tilde{\mathcal{P}}$ that employs $2\log_2d$ single qubit exchanges.
One can achieve this by splitting a $q$-qubit message from Alice to Bob (or the other way round) into $q$ rounds of one qubit exchanges. The new protocol $\tilde{\mathcal{P}}$ has a total of $\tilde{R}=\log_2d -1$ rounds, with each round involving transmission of a qubit from Alice to Bob and back. In the last round Alice sends a qubit $\tilde{\alpha}_{\tilde{R}}$ and Bob instead of sending back one, measures using another measurement $\{\tilde{M}^y_z\}$ the joint system of her local memory $\tilde{B}^{x,y}_{\tilde{R}-1}$ and the qubit message from Alice $\tilde{\alpha}_{\tilde{R}}$. The winning probability for $\tilde{\mathcal{P}}$ is equal to success probability of $\mathcal{P}$ but has the expression 
\be \label{pQd2}
p_{\mathcal{Q}_d}=\sum_{x,y}p(x,y)\tr(\rho_{\tilde{\alpha}_{\tilde{R}},\tilde{B}^{x,y}_{\tilde{R}-1}}\tilde{M}^y_z).\ee

\item In the protocol $\tilde{\mathcal{P}}$, in each round $r$ Bob applies a unitary $\tilde{U}^y_r$ on the one qubit message from Alice from the previous round $\tilde{\alpha}_{r-1}$ and her local memory $\tilde{B}^{x,y}_{r-1}$. One can view the unitary operation as a controlled gate acting on the memory with one qubit message being the control. This observation implies that for a fixed input $x$, for round $r$ (i.e. after $r-1$ rounds), Bob's memory is spanned on at-most $2^{r-1}$ orthogonal vectors (see \textit{Lemma 2} in \citep{pnas}). This implies that for the last round Bob's memory in $\tilde{\mathcal{P}}$ requires at-most $\tilde{R}-1$ qubits and the state $\rho_{\tilde{\alpha}_{\tilde{R}},\tilde{B}^{x,y}_{\tilde{R}-1}}$ is at-most $d$-dimensional (or equivalently $\log_2d$-qubits).
\end{enumerate} 
Now we are prepared to present our result concerning $\mathcal{PC}$ manifest in advantageous two-way multi-round quantum $\mathcal{CC}$ protocols,
\begin{result}
Given a two-way multi-round quantum $\mathcal{CC}$ protocol $\mathcal{P}$, an advantage is obtained in the $\mathcal{OC}$ task described in \eqref{oc}  with $p(a_2=0|x)=1/d^{n_y}$ ($p_\mathcal{Q}>  p_\mathcal{NC}$)  whenever the following condition holds,
\be \label{c77}
\frac{1}{d^{n_y}}(2p_{\mathcal{Q}_d} + d^{n_y}-1 - d^{n_y-1}\chi) > p_{\mathcal{C}_2}.
\ee
where $\chi = \sum_{x,y} p(x,y)\tr\left( \tilde{M}^y_{z=f(x,y)}\right)$ and $\{\tilde{M}^y_{z}\}$ are Bob's measurements employed in the derived quantum $\mathcal{CC}$ protocol $\tilde{\mathcal{P}}$.
\end{result}
\begin{proof} We begin by devising a quantum strategy for the $\mathcal{OC}$ task. 
We orchestrate a quantum strategy for the $\mathcal{OC}$ task based on the quantum two-way multi-round $\mathcal{CC}$ protocol. To achieve this we exploit the fact that there is no-restriction on the amount of communication in the $\mathcal{OC}$ task.
The core idea remains the same as in one-way $\mathcal{CC}$ case, Alice sends Bob's pre-measurement state when $a_2=0$ and its orthogonal mixture when $a_2=1$. We start with converting the given quantum two-way multi-round $\mathcal{CC}$ protocol $\mathcal{P}$ which uses $d$-dimensional communication in total, to one that uses $2\log_2d$ qubits of communication $\tilde{\mathcal{P}}$. There is still an issue with this approach, Alice does not know the value $y$ in advance, and the pre-measurement state may depend on $y$. In order to deal with this issue, when $a_2=0$ Alice simply prepares the pre-measurement states for all values of $y$ and sends a tensor product of these states as the message $\Theta_{x,a_2=0}=\bigotimes_y\rho_{\tilde{\alpha}_{\tilde{R}},\tilde{B}^{x,y}_{\tilde{R}-1}}$. Recall that the states $\rho_{\tilde{\alpha}_{\tilde{R}},\tilde{B}^{x,y}_{\tilde{R}-1}}$ are at-most $d$-dimensional. 
When $a_2=1$, Alice sends the orthogonal mixture of $\Theta_{x,a_2=0}$, $\Theta_{x,a_2=1} = \frac{\mathbb{I}-\bigotimes_y\rho_{\tilde{\alpha}_{\tilde{R}},\tilde{B}^{x,y}_{\tilde{R}-1}}}{d^{n_y}-1}$. It is straightforward to see that Alice's preparation are oblivious to $x$, as $\forall x,x'\in[n_x], \ \sum_{a_2}p(a_2,x)\Theta_{x,a_2}=\sum_{a_2}p(a_2,x')\Theta_{x',a_2}={\frac{\mathbb{I}}{d^{n_y}}}$.
  Now, upon receiving the message from Alice, Bob performs the measurement $\tilde{M}^y_z$ on the relevant part (depending on his input $y$) of the message i.e. either $\rho_{\tilde{\alpha}_{\tilde{R}},\tilde{B}^{x,y}_{\tilde{R}-1}}$ or  $tr_{\lnot y}(\Theta_{x,a_2=1})=\frac{d^{n_y-1}\mathbb{I}-\rho_{\tilde{\alpha}_{\tilde{R}},\tilde{B}^{x,y}_{\tilde{R}-1}}}{d^{n_y}-1}$.
This strategy yields the following success probability,
\begin{align} \label{myLove1}
p_{\mathcal{Q}} & = \sum_{x,a_2=0,y} p(x,a_2,y) \tr(\rho_{\tilde{\alpha}_{\tilde{R}},\tilde{B}^{x,y}_{\tilde{R}-1}} \tilde{M}^y_{z=f(x,y)}) \nonumber \\ \nonumber & \;\;\;\;\;\;\;\; + \sum_{x,a_2=1,y} p(x,a_2,y) \tr\left(\frac{d^{n_y-1}\mathbb{I}-\rho_{\tilde{\alpha}_{\tilde{R}},\tilde{B}^{x,y}_{\tilde{R}-1}}}{d^{n_y}-1}\tilde{M}^y_{z=1\oplus f(x,y)} \right) \nonumber \\  & = \frac{1}{d^{n_y}}(2p_{\mathcal{C}_d} + d^{n_y}-1 - d^{n_y-1}\chi), 
\end{align} where $\chi = \sum_{x,y} p(x,y)\tr\left( \tilde{M}^y_{z=f(x,y)}\right)$ and $p_{\mathcal{Q}_d}$ is given by \eqref{pQd2}.  Now our desired result simply follows from \eqref{PNCineq1}.
\end{proof}
Given quantum advantage in $\mathcal{CC}$ problem ($p_{\mathcal{Q}_d}>p_{\mathcal{C}_d}$) and \eqref{myLove1}, an advantage is obtained in the $\mathcal{OC}$ task ($p_\mathcal{Q}>  p_\mathcal{NC}$) described in \eqref{oc} with $p(a_2=0|x)=1/d^{n_y}$ whenever the following holds,
\be \nonumber
\frac{1}{d^{n_y}}(2p_{\mathcal{C}_d} + d^{n_y}-1 - d^{n_y-1}\chi) > p_{\mathcal{C}_2}  .
\ee


\subsubsection{Entanglement assisted classical communication protocols}
Another non-equivalent \cite{armin,us} class of advantageous quantum $\mathcal{CC}$ protocols is that of entanglement assisted classical communication protocols, wherein Alice and Bob share an entangled state $\rho_{AB}$ (a density operator on $\mathcal{H}_A\otimes \mathcal{H}_B$), Alice performs a $d$ outcome measurement ($\{M^x_m\}$) and sends her outcome $m$ as the message. Upon receiving the message $m$, Bob performs a binary outcome measurement ($\{M^{y,m}_z\}$). The quantum guessing probability is expressed as
\be \label{pQd3}
p_{\mathcal{Q}_d} = \sum_{x,y} p(x,y) \sum^{d-1}_{ m=0}\tr(\rho_{AB}M^x_{m}\otimes M^{y,m}_{z=f(x,y)}).\ee  Let the reduced density matrix of Bob's part of the entangled state $\rho_{B}$ be of dimension $e$ i.e. $e=dim(\mathcal{H}_{B})$. A quantum strategy for the $\mathcal{OC}$ task \eqref{oc} based on advantageous entanglement assisted classical communication $\mathcal{CC}$ protocols and the corresponding condition for retrieving an advantage is presented in the following result,
\begin{result}
Given a entanglement assisted  classical communication $\mathcal{CC}$ protocol, an advantage is obtained in the $\mathcal{OC}$ task described in \eqref{oc} with $p(a_2=0|x)=1/d'$ ($p_\mathcal{Q}>  p_\mathcal{NC}$)  whenever the following condition holds,
\be \label{c777}
\frac{1}{d'}(2p_{\mathcal{Q}_d} + d'-1 - \chi) > p_{\mathcal{C}_2}.
\ee
where $\chi = \sum_{x,y} p(x,y)\tr\left( {M}^y_{z=f(x,y)}\right)$ and $\{{M}^y_{z}\}$ are Bob's measurements employed in the $\mathcal{CC}$ protocol, $d'=de$ and $e$ is the dimension of Bob's local part of the shared entangled state.
\end{result}
\begin{proof}
In this case, we capitalize over the fact that the amount of communication is unrestricted in the $\mathcal{OC}$ task and convert the given entanglement assisted classical communication protocol to a prepare and measure protocol wherein Alice simply sends Bob the corresponding pre-measurement state (Bob's marginal state along with the classical message). This in-turn enables us to construct quantum strategies for the $\mathcal{OC}$ task employing the aforementioned methodology.\\
In order to utilize the machinery developed so far we first construct a quantum prepare and measure protocol deploying a $d'=de$ dimensional communicated system but with the same probability of success $p_{\mathcal{Q}_d}$ as the given entanglement assisted classical communication protocol. Upon receiving $x$ Alice prepares the state $\rho_x=|m\rangle \langle m| \otimes \rho_{B}$ where the state $|m\rangle \langle m|$ is simply the quantum encoding of the classical message $m$ into $d$ orthogonal states. She accomplishes this feat by measuring $\{M^x_m\otimes \mathbb{I}\}$ on the entangled state $\rho_{AB}$ to which we assume she has access to. The communicated system is of dimension $d'=de$. Bob first retrieves the message by performing the measurement $\{M_m\}$ on the appropriate subsystem of the communicated system and depending on it performs the measurement $\{M^{y,m}_z\}$ on rest of the communicated system, captured conveniently in a joint measurement $\{\tilde{M}^y_z=M_m\otimes M^{y,m}_{z}\}$. This yields the same success probability $p_{\mathcal{Q}_d}$. Now, we convert this prepare and measure protocol into an $\mathcal{OC}$ protocol utilizing the methodology described in the proof of \textbf{Result 1} and obtain the following lower bound on quantum success probability for the $\mathcal{OC}$ task,
$p_\mathcal{Q}\geq \frac{1}{d'}\left( 2p_{\mathcal{Q}_d} + d'-1 - \chi \right)$ where $\chi= \sum_{x,y} p(x,y)\tr\left( M^{y}_{z=f(x,y)}\right)$ and $p_{\mathcal{Q}_d}$ is given in \eqref{pQd3}. 
This in-turn leads us to the condition for quantum advantage in the $\mathcal{OC}$ task \eqref{c777}.
\end{proof}
Notice that in a rather predominant subclass of entanglement assisted classical communication protocols Bob applies a completely-positive trace preserving map $\Lambda_m$ on his part of the entangled state $\rho_\mathcal{B}$ and performs the measurement $\{M^y_z\}$ on $\Lambda_m(\rho_{B})$. In such cases Alice having access to the message $m$ sends $\rho_x=\Lambda_m(\rho_{B})$ effectively 
reducing the dimension of the communicated system in the prepare measure protocol to $d'=e$, thereby improving the feasibility of the quantum advantage in the $\mathcal{OC}$ task.\\

 \subsection{Unbounded violation of $\mathcal{PNC}$ inequalities} 
 To demonstrate the vitality of the results obtained so far we illustrate two examples of unbounded quantum violations of $\mathcal{PNC}$ inequalities based of two widely studied $\mathcal{CC}$ problems and associated prepare and measure protocols with exponential quantum advantage.
Let us re-write the $\mathcal{PNC}$ inequality \eqref{PNCineq1} as
$\alpha_{{\mathcal{NC}}}\leq \alpha_{\mathcal{C}_2},$
where $\alpha_{\mathcal{NC}}=p_{\mathcal{NC}}-\frac{1}{2},\alpha_{\mathcal{C}_2}=p_{\mathcal{C}_2}- \frac{1}{2}$. Then a quantum advantage in a $\mathcal{CC}$ problem adhering to the condition \eqref{c7} implies that there exists quantum protocol for the $\mathcal{OC}$ task with $\alpha_{\mathcal{Q}}=\frac{1}{d}(2p_{\mathcal{Q}_d} + d-1 - \chi) - \frac{1}{2}$. Quantum advantage in $\mathcal{CC}$ problems is prevalently reported in terms of the amount of communication required to achieve a bounded probability of success $p_{\mathcal{S}}$, i.e., $\mathcal{Q}(f,p_{\mathcal{S}})<\mathcal{C}(f,p_{\mathcal{S}})$. To apply our results to the innumerable instances of quantum advantage reported in this fashion, we employ the following lemma,
\begin{lemma} \label{Lemma5} Given a $\mathcal{CC}$ problem and a protocol which achieves a success probability $p_{S}$ using $\mathcal{C}(f,p_\mathcal{S})$ bits, the success probability of a protocol using a two-leveled classical message is upper bounded in the following way,
\be \label{pump2}
p_{\mathcal{C}_2}\leq \frac{1}{2}  + \sqrt{\frac{2p_S}{\mathcal{C}(f,p_S)}}.
\ee
\end{lemma}
The proof has been deferred to the Appendix. 
\begin{fact} \label{coro3}
The ratio of quantum and $\mathcal{PNC}$ values of $\alpha$ (denoted by $\beta$) can be lower bounded with help of \textcolor{blue}{\textit{Lemma}} \ref{le:3} in the following way,
\beq \label{beta}
& \beta & \geq \frac{\alpha_{\mathcal{Q}}}{\alpha_{\mathcal{NC}}} \geq \frac{\frac{1}{d}(2p_{\mathcal{Q}_d} + d-1 - \chi)-\frac{1}{2}}{p_{C_2} - \frac{1}{2}} \nonumber \\
&& \geq \frac{\sqrt{\mathcal{C}(f,p_S)}(2p_{\mathcal{Q}_d} + d/2 -dp_G -1)}{d\sqrt{2p_S}}.
\eeq 
\end{fact}
To obtain an unbounded violation of the $\mathcal{PNC}$ inequality $\alpha_{{\mathcal{NC}}}\leq \alpha_{\mathcal{C}_2},$ it suffices to show that $\beta$ could be arbitrarily large ($>> 1$) \cite{uv}. We demonstrate the same for two widely studied $\mathcal{CC}$ problems \cite{vsp,hm} with exponential quantum advantage, \\
1. \textit{Vector in a subspace:} Alice is given an $n$-dimensional unit vector $u$
and Bob is given a subspace of dimension $n/2$, $S$ with the promise that either
$u \in S$ or $u \in S^\perp$. Their goal is to decide which is the case. Here $p_{\mathcal{Q}_d=\log n} =1$ i.e. $\mathcal{Q}(f,1)=\log n,$ $\mathcal{C}(f,p_S=\frac{2}{3}) = \Omega (\sqrt[3]{n})$ (Theorem 4.2 in \cite{vsp}) and a simple calculation yields $\chi = \frac{\log n}{2}, p_G = \frac{1}{2}$. Inserting these into \eqref{beta} one obtains an arbitrarily large lower bound for the ratio 
$ \beta \geq \Omega(\frac{\sqrt[6]{n}}{\log n}) .$ \\
2. \textit{Hidden matching: } Alice is given a bit string $x \in \{0,1\}^n$ of length $n$ and Bob is given $y \in M_n$ ($M_n$ denotes the family of all
possible perfect matchings on $n$ nodes). Their goal is to output a tuple $( i, j, t )$ such
that the edge $(i, j)$ belongs to the matching $y$ and $t=x_i \oplus x_j$. Clearly the hidden matching problem is not a typical $\mathcal{CC}$ problem, specifically it is a relational problem. Nevertheless, we can find that the machinery developed so far including the \textit{Proposition} \ref{prop:pncIneqMain} and \textit{Corollary} \ref{coro3} still hold for relational $\mathcal{CC}$ problem.
\begin{lemma} \label{le:hm} For Hidden matching problem an $\mathcal{OC}$ task can be constructed with a success probability $p_{\mathcal{NC}}$, such that $p_\mathcal{NC} \leq p_{\mathcal{C}_2}$.
\end{lemma}
The proof is similar to the proof of \textit{Proposition} \ref{prop:pncIneqMain} (see  Appendix). This proof provides for our insight that our results persist in case of relational $\mathcal{CC}$ problems beyond main-stream functional $\mathcal{CC}$ problems. For Hidden matching $p_{\mathcal{Q}_d} =1, \mathcal{Q}(f,1)=d=\log n, p_G = \frac{1}{2}$, $\chi = \frac{\log n}{2}$ and $\mathcal{C}(f,1) = \Omega (\sqrt[]{n})$ \cite{hm}. Inserting these observations into \eqref{beta} one obtains an even larger violation as the lower bound on $\beta$ grows faster, i.e., 
$\beta \geq \Omega(\frac{\sqrt[4]{n}}{\log n})$.

\subsection{Alternative construction of $\mathcal{OC}$ task}
\begin{figure}[http]
\centering
\includegraphics[scale=0.67]{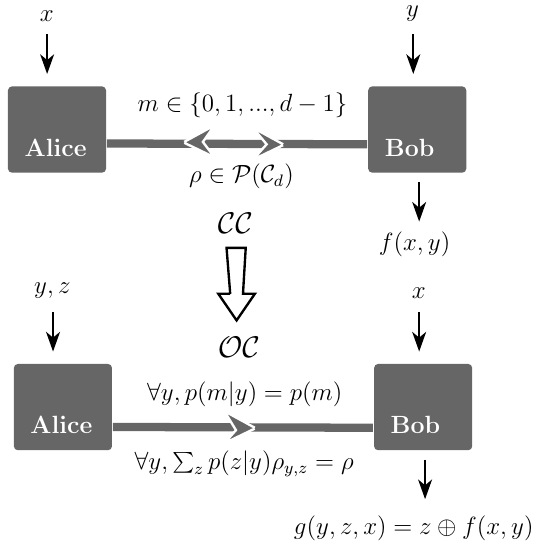}
\caption{Alternative Construction of $\mathcal{OC}$ task based on a given $\mathcal{CC}$ task. The communication is restricted so as not to reveal any information about a oblivious variable $y$.}
\label{fig3}
\end{figure}

An equivalent alternative construction of the $\mathcal{OC}$ task tailored to a given $\mathcal{CC}$ problem is presented here. 
Given a general $\mathcal{CC}$ problem and an advantageous quantum $\mathcal{CC}$ protocol, i.e., $p_{\mathcal{Q}_d} > p_{\mathcal{C}_d}$, we construct the following $\mathcal{OC}$ task (shown is Fig. \ref{fig3}),
\beq  \label{oc2}
&a = (y,z), \ b = x,\ c \in \{0,1\}, \nonumber \\
&p(a,b) = p(y,z,x) = p(x)p(y|x)p(z|y), \nonumber \\
&\text{where } p(z|y) = \frac{\tr(M^y_z)}{d}, \nonumber \\
& g(y,z,x) = f(x,y) \oplus z . 
\eeq
 Here $\{{M}^y_{z}\}$ are Bob's measurements employed in the given quantum $\mathcal{CC}$ protocol under consideration, and the oblivious condition constraints the communicated system to not carry any information about y.

\begin{proposition} \label{propo2} 
The $\mathcal{PNC}$ success probability of the $\mathcal{OC}$ task described in \eqref{oc2} is upper bounded by the maximal classical success probability of the $\mathcal{CC}$ problem wherein Alice is restricted to communicate a two-leveled system, i.e. $p_\mathcal{NC} \leq p_{\mathcal{C}_2}$. 
\end{proposition}
\begin{proof} We follow the same steps as in the proof of \textit{Proposition} \ref{prop:pncIneqMain}. Again employing \textit{Lemmas} \ref{le:UpperBound} and \ref{le:extremal} we arrive at the following upper on the classical success probability of the $\mathcal{OC}$ task described in \eqref{oc2}, 
\beq \label{A2}
&p_{\mathcal{NC}} \leq & \sum_x p(x) \sum_{y,z} p(y|x) \ \delta_{e_{y},z} \ \delta_{c_x,z\oplus f(x,y)} \nonumber \\
&& = \sum_{x,y} p(x,y) \ \delta_{e_{y},c_x\oplus f(x,y)}\ .
\eeq
Let's consider the following classical protocol employing a two-leveled message $m\in \{0,1\}$ for the $\mathcal{CC}$ problem,
\begin{align}
p_{E}(m|x) = \delta_{m,c_x} \ , \ p_{D}(z|y,m) =   \delta_{z,m \oplus e_y}. 
\end{align}
Inserting the above strategy in \eqref{PCOWCC1},  one obtains the same success probability in $\mathcal{CC}$ problem as given in the right side of \eqref{A2}. 
\end{proof}

The contrasting feature of this construction is that the exact duals of the states and measurements used in the advantageous quantum $\mathcal{CC}$ protocol form the corresponding measurements and states respectively for the quantum $\mathcal{OC}$ protocol. That is, Alice's preparation for the $\mathcal{OC}$ task are $\rho_{y,z} = \frac{M_z^y}{\tr(M^y_z)}$ and Bob's measurement 
for his input $x$ is $\{\rho_x,\mathbb{I}-\rho_x\}$. Clearly Bob remains oblivious to $y$ due to the completeness of quantum measurements i.e. $\forall y, \ \sum_{z} p(z|y) \rho_{y,z}=\frac{\mathbb{I}}{d}$. Subsequently, plugging the expressions of $p(y,z,x)$ from \eqref{oc2} and $p_{\mathcal{Q}_d}$ from \eqref{pQd1}, a simple calculation leads to the same expression as in  \eqref{myLove},
\begin{align}
p_{\mathcal{Q}} & = \sum_{y,z=f(x,y),x} p(y,z,x) \tr\left(\rho_x \frac{M^y_{z}}{\tr(M^y_{z})} \right) + \nonumber \\ & \hspace{10pt}  \sum_{y,z=1\oplus f(x,y),x} p(y,z,x) \tr\left( (\mathbb{I} - \rho_x)\frac{M^y_{z}}{\tr(M^y_{z})} \right) \nonumber \\
& = \frac{1}{d}\left( 2p_{\mathcal{Q}_d} + d-1 - \chi \right)
\end{align}
where $\chi=\sum_{x,y} p(x,y)\tr\left( M^y_{z=f(x,y)}\right)$. Thus, all the results derived previously remain intact for this alternative construction of $\mathcal{OC}$ task. \\

This construction provides for our inference that our main results can be extended to general probabilistic theories with the feature of self-duality of states and measurement effects \cite{book,carlo}. This follows from the fact that the states and measurements that reveal $\mathcal{PC}$ in the alternative $\mathcal{OC}$ task are just the dual of the measurement effects and states employed in the $\mathcal{CC}$.
The property of self-duality emerges from a set of natural postulates in the framework of general probabilistic theories \cite{carlo}. 
However, this implication is not true in any operational theory. Here, we demonstrate a toy-theory and an ontic-model with $\mathcal{CC}$ advantage but no possibility of $\mathcal{PC}$. Consider a well-known $\mathcal{CC}$ task, the $(2 \rightarrow 1)$ random access code \cite{rac} wherein Alice receives two random input bits $x_1,x_2$ to be encoded into a two dimensional system and sends it to Bob. Bob receives a random input bit $y$ along with the message from Alice and is required to guess $x_y$. This theory, having only three preparations and just two measurements, is a fragment of quantum theory. This fragment of quantum theory doesn't adhere to self-duality. Clearly the theory admits advantage in this task as the average success probability $p_{\mathcal{Q}_2} \approx 0.8 > p_{\mathcal{C}_2}=0.75$. In an ontological model underlying this toy theory there are only be three ontic states labeled as $\psi_{x_1x_2}$ which correspond to pure quantum preparations as, $\psi_{11,10} = |1\rangle, \psi_{00} = \cos(\theta) |0\rangle + \sin(\theta) |1\rangle,\psi_{01} =  \cos(\theta) |0\rangle - \sin(\theta) |1\rangle$ where $\theta = \frac{\pi}{8}$ and two binary-outcome response schemes corresponding to Bob's setting $y=0,1$ and measurements $\sigma_z, \sigma_x$ respectively.  However since this ontological model has only three ontic states, any mixed preparation in this theory has a unique decomposition, thus ruling out the possibility of $\mathcal{PC}$ \cite{Pusey}. This shows from the basis of the inference that self-duality of states and measurements is a necessary requirement for our results to persist in general probabilistic theories.

\section{Bell inequality violation implies advantage in $\mathcal{OC}$}
With the help of the tools developed so far we now present the complete proof of the fact that Bell inequality violations imply advantage in an associated $\mathcal{OC}$ task.
For any Bell inequality an $\mathcal{OC}$ task can be constructed porting Bell-inequality violation to an advantageous strategy for the $\mathcal{OC}$ task. For the space-like separated scenario the collapsed state on Bob's end is prepared and sent in the $\mathcal{OC}$ task and for the time-like separated case \cite{LP} the pre-measurement state at Bob's end is prepared and sent in the $\mathcal{OC}$ task. This would make all Bell-inequality violation operationally reveal $\mathcal{PC}$. However, there is a subtlety here, while deterministic encoding strategies yield bounds on Bell inequalities, the $\mathcal{PNC}$ bounds on the success parameter of the $\mathcal{OC}$ tasks might spring from probabilistic encoding schemes \cite{shp}. An inadequate attempt to prove the above thesis was made in \cite{chutiyapa}, as the authors explicitly assume deterministic encoding schemes for the constructed $\mathcal{OC}$ task. We use the tools developed in this article to provide the complete proof for the thesis. \\
The set-up for a space-like separated Bell experiment does not involve any communication, instead two spatially separated parties Alice and Bob are provided with inputs $x\in [n_x],y\in [n_y]$ respectively. Their objective is to return outputs $u\in [n_u],v\in [n_v]$ respectively so as to maximize an expression of the following form,
\be 
\mathcal{B} = \sum_{u,v,x,y} s_{x,y,u,v} \ p(x,y) p(u,v|x,y),
\ee
where $s_{x,y,u,v} \geq 0$. The parties may share correlations (classical: shared randomness or quantum: entangled states) which essentially yield advice in the form of conditional probability distributions $p(u,v|x,y)$. If Alice and Bob share a local-realist (classical) correlation, the maximum they can achieve is,
\be  \label{bl}
\mathcal{B}_\mathcal{L} = \sum_{\lambda,u,v,x,y} s_{x,y,u,v}\ p(x,y) p(\lambda) p_{\lambda}(u|x) p_{\lambda}(v|y).
\ee 
This fact is captured in Bell inequalities. \\
Consider a quantum strategy which violates a Bell inequality i.e. $\mathcal{B}_\mathcal{Q} > \mathcal{B}_{\mathcal{L}}$. The probability of getting outcome $u$ when measurement $x$ is performed on the shared quantum state is $p_\mathcal{Q}(u|x)$ and the reduced quantum state on Bob's subsystem is denoted by $\rho^B_{u|x}$. We follow the construction of $\mathcal{OC}$ presented in \cite{chutiyapa}, 
\beq
&a=(a_1,a_2) = (x,u), \ b = y, \ c=v , \nonumber \\
&p(a,b) = p(x,u,y) = p(y)p(x|y)p_\mathcal{Q}(u|x),
\eeq
where communication is constrained to oblivious to $x$.
The figure of merit in the $\mathcal{OC}$ is given by,  
\be \label{pbi}
p = \sum_{u,v,x,y} s_{x,y,u,v}\ p(x,u,y) p(v|x,u,y).\ee

\begin{proposition}
The non-contextual success probability of the $\mathcal{OC}$ task is upper bounded by the optimal local-realist value of Bell expression, i.e. $p_\mathcal{NC} \leq \mathcal{B}_\mathcal{L}$. 
\end{proposition}  
\begin{proof}
It is straightforward to see that \textit{Lemmas} \ref{le:UpperBound} and \ref{le:extremal} apply just as well to the above $\mathcal{OC}$ task and similar to \eqref{optnc*} we retrieve an upper-bound on the associated $p_{\mathcal{NC}}$ as follows,
\beq \label{A3}
p_\mathcal{NC} & \leq &\sum_{y} p(y) \sum_{x,u,v} p(x|y) s_{x,y,u,v}\ \delta_{e_x,u} \ \delta_{v_y,v} \nonumber \\
&=& \sum_{x,y,u,v} p(x,y) s_{x,y,u,v}\ \delta_{e_x,u} \ \delta_{v_y,v} \ .
\eeq
Now, we detail the proof of the above observation.
The expression for maximal classical success probability \eqref{pbi} is,
\beq \nonumber
&p_{\mathcal{NC}} = & \max_{\{\mathcal{E}\}\{\mathcal{D}\}} \bigg\{ \sum_{m} \sum_{y} p(y) \\ & &
  \bigg(\sum_{x,u} p(x|y)p_\mathcal{Q}(u|x) s_{x,y,u,v}\ p_{\mathcal{E}}(m|x,u) p_{\mathcal{D}}(v|y,m) \bigg)\bigg\} , \nonumber \\
\eeq
and the oblivious constraints imply,
\beq \nonumber
&\forall m,\forall x,x'\in [n_x], \\ \nonumber & \quad p_{\mathcal{E}}(m) &:= p_{\mathcal{E}}(m|x) \\ \nonumber & &= \sum_{u} p_{\mathcal{Q}}(u|x) p_{\mathcal{E}}(m|x,u) \\  & &= p_{\mathcal{E}}(m|x') .
\eeq
Now following the same argument as in the proof of \textit{Lemma} \ref{le:UpperBound} one obtains,

\begin{align} \label{optsnc}
&p_{\mathcal{NC}} \leq \max_{\{q_{x,u}\}} \bigg\{ \sum_y p(y) \max_v \bigg\{
\sum_{x,u} p(x|y) q_{x,u} s_{x,y,u,v} \bigg\} \bigg\},  \\
&\text{where} \quad \forall x,u, \ q_{x,u}\geq 0, \ \sum_u q_{x,u}=1. \nonumber
\end{align}
Now invoking \textit{Lemma} \ref{le:extremal}, suppose the extremal point yielding the optimal value of right-hand-side of \eqref{optsnc} corresponds to $u^{ext}=e_{x}$ for each $x$, i.e., $q_{x,u}=\delta_{u,e_{x}}$,  and for that extremal point 
$\max\limits_v \{
\sum_{x,u} p(x|y) q_{x,u} s_{x,y,u,v} \}$ is achieved for $v_y$ for each $y$. Subsequently, \eqref{optsnc} can be expressed as \eqref{A3}. 

Now we propose a hidden variable model such that
$p_{\lambda}(u|x) = \delta_{u,e_x}\ , \ p_{\lambda}(v|y) = \delta_{v,v_y} $.
Plugging this local strategy into \eqref{bl}, one obtains the same the expression for $\mathcal{B_L}$ as the right-hand-side of \eqref{A3}, thus completing the proof.  
\end{proof}

A quantum strategy for the $\mathcal{OC}$ task can be easily constructed from the states and measurements responsible for violation of Bell inequality: Alice sends $\rho^B_{u|x}$ for input $(x,u)$ and Bob's measurement settings are the same as in the given Bell experiment. Adherence of oblivious condition for this strategy simply follows from the no-signaling condition. Thus, we conclude $p_\mathcal{Q} = \mathcal{B}_\mathcal{Q} > \mathcal{B}_\mathcal{L} \geq p_\mathcal{NC}$.

\section{Conceptual insight and implications} 
The early stages of the quantum information epoch focused primarily on finding communication, computation and information processing tasks wherein quantum resources and protocols provide advantage over their classical counterparts. As a consequence, the quantum departure from classical limits in such tasks has been significantly substantiated in innumerable and variegated classes of tasks, this perception is now commonly referred to as the "quantum advantage". However, there is little insight into what feature of quantum theory is underneath such a remarkable feat. Consequently, further search for such tasks usually employs narrowing heuristic intuition. The answers to such questions carry with them the potential of directing and broadening the search for tasks with quantum advantage. However, this seemingly simple question turns out to be substantially arduous and rich in complexity. We must begin by discarding the cyclic answers that inherently refer back to the operational quantum formalism which was apriori responsible for the advantageous predictions such as superposition of states. While these answers might lead to sharpening intuition, they don't lead to any significant insights. To further insight, the answers must arguably pertain to a ground common to classical and quantum theory, where non-classical features underlying the quantum formalism are uncovered. The study of ontology or "underlying hidden variables" provides for such a ground. On the other hand, quantum communication advantage has a vast variety of manifestations, however, quantum $\mathcal{CC}$ advantage and device-independent information processing form the most prominent of them. In this article, we sought to find the quantum ontic-feature that underlies quantum $\mathcal{CC}$ advantage. \\
In a nutshell, this work exposes the essential connection between operational quantum communication advantage and the ontic-feature of $\mathcal{PC}$, via operational $\mathcal{OC}$ tasks. In other words, we unveil a unifying connection between quantum $\mathcal{CC}$ advantage and quantum advantage in $\mathcal{OC}$ tasks, where the later forms the operational signature of $\mathcal{PC}$.
We provide two intuitive ways of constructing an $\mathcal{OC}$ task tailored to any given $\mathcal{CC}$ task \eqref{oc} and \eqref{oc2}. The $\mathcal{OC}$ tasks thus obtained have two salient features: First, the maximal achievable classical success probability in both $\mathcal{OC}$ tasks is bounded by the maximal achievable classical success probability in the $\mathcal{CC}$ problem when the communicated system is restricted to be two-dimensional. This in-turn provides for two-distinct $\mathcal{PNC}$ inequalities corresponding to every $\mathcal{CC}$ problem.
Second: for any advantageous quantum $(i)$ prepare and measure, $(iii)$ two-way multi-round and, $(iii)$ entanglement assisted classical communication $\mathcal{CC}$ protocols, we obtain quantum $\mathcal{OC}$ strategies which utilize the same states and measurements. An advantage is obtained in the constructed $\mathcal{OC}$ task revealing $\mathcal{PC}$ whenever the conditions \eqref{c7},\eqref{c77} and \eqref{c777} are met respectively. It is a remarkable accomplishment of our construction, that these conditions feature a comparison between $\mathcal{CC}$ performance of quantum $d$-level and $\mathcal{CC}$ performance of classical $2$-level systems. Notably, these conditions allow us to demonstrate first instances of unbounded violation of $\mathcal{PNC}$ inequalities from exponential quantum $\mathcal{CC}$ advantage.
We remark that there exists a trade-off between generality of our results and the tightness of these conditions for higher dimensional quantum $\mathcal{CC}$ protocols. Because in this work, we concern ourselves with general implications, these already substantially tight conditions might be tightened even further by fine-tuning our constructions to specific $\mathcal{CC}$ problems and associated higher dimensional quantum $\mathcal{CC}$ protocols. For instance, these conditions base themselves on the $\mathcal{PNC}$ inequality in \textit{Proposition} \ref{prop:pncIneqMain} which in-turn relies on a state of the art technique we employed to obtain upper bounds on maximal classical ($\mathcal{PNC}$) success probability of $\mathcal{OC}$ tasks. A tighter upper bound on maximal classical success probability $p_\mathcal{NC}$ or finding out the exact value will further tighten the conditions under consideration.
In summary, not only do our results capture $\mathcal{PC}$ manifest in all predominant classes of advantageous quantum $\mathcal{CC}$ protocols but they also hold beyond mainstream functional $\mathcal{CC}$ problems i.e. even in case of relational $\mathcal{CC}$ problems (see proof of \textit{Lemma} \ref{le:hm}). 
 \\
Our two constructions underscore two distinct ways in which $\mathcal{PC}$ is manifest in an advantageous $\mathcal{CC}$ protocol.
An advantage in the first $\mathcal{OC}$ task \eqref{oc} {reveals $\mathcal{PC}$ manifest} in the states from the $\mathcal{CC}$ protocol and their orthogonal mixtures, using the same measurements from the $\mathcal{CC}$ protocol. Whereas an advantage in the second $\mathcal{OC}$ task \eqref{oc2} reveals $\mathcal{PC}$ manifest in the states corresponding to the measurement effects from the $\mathcal{CC}$ protocol, with the aid of measurements corresponding to the states employed in the $\mathcal{CC}$ task. The second construction \eqref{oc2} enables a direct inference that our results and implications can be extended beyond quantum theory in general probabilistic theories with the property of self-duality of states and measurements effects.\\
Concerning other ontic-features as plausible ground of quantum $\mathcal{CC}$ advantage, the connection between quantum advantage in $\mathcal{CC}$ and non-locality has been explored in \cite{BZJZ,pnas,SL}. Given any protocol offering a sufficiently large quantum $\mathcal{CC}$ advantage, \cite{pnas,SL} provide a way for obtaining measurement statistics that violate some Bell inequality. These approaches basically employ an independent teleportation subroutine to transmit Alice's preparations (from quantum $\mathcal{CC}$ protocol). This in-turn implies that the non-locality thus revealed stems from additional entangled states and measurements associated with the teleportation protocol, which are unrelated to the ones employed in the advantageous quantum $\mathcal{CC}$ protocol. Therefore, the assertion that quantum $\mathcal{CC}$ advantage implies non-locality is rather weak. 
Whereas, along with the very states and measurements responsible for the quantum $\mathcal{CC}$ advantage we use additional preparations, but these preparations are orthogonal mixtures of these states and therefore depend on the advantageous protocol. Therefore in this sense, our results reveal a substantially more intimate connection between quantum $\mathcal{CC}$ advantage and $\mathcal{PC}$.  Furthermore, we provide a complete proof of the fact that any Bell-inequality violation implies an advantage in an associated $\mathcal{OC}$ task, thereby porting even the weak implication along with device-independent information processing 
operationally to $\mathcal{PC}$. Moreover, \cite{shp} shows that all logical proofs of Kochen-Specker contextuality yields an advantage in the $\mathcal{OC}$ task. It is a well-known fact that while a two dimensional quantum system is enough to demonstrate $\mathcal{PC}$, Kochen-Specker contextuality and non-locality require at-least three and four dimensional quantum systems respectively. In summary not only a wide-spectrum of quantum communication advantage reveals $\mathcal{PC}$, even the operational witnesses of other well known ontic-features imply $\mathcal{PC}$. 
This leads us to our tentative assertion that $\mathcal{PC}$ is inmately related to quantum communication advantage. \\
While our implications are ontological, our methodology is strictly operational and employs advantage in $\mathcal{OC}$ tasks as the intermediary between operational $\mathcal{CC}$ advantage and the ontic-feature of $\mathcal{PC}$. Our results therefore indicate the fundamental significance of $\mathcal{OC}$ tasks to quantum advantage in communication. Furthermore, $\mathcal{OC}$ tasks form primitives for a range of cryptographic protocols \cite{crypto,qcrypto} and have found applications in privacy-preserving computation \cite{ppc}. Apart from the aforementioned implications our methodology has exposed a large class of $\mathcal{OC}$ tasks with quantum advantage.

The question ``why quantum advantage?" is far from settled. While the results of this article point to $\mathcal{PC}$, they in no-way close the door to more fundamental ontological or causal features of quantum theory.
A much more arduous question of whether $\mathcal{PC}$ with self-duality (or some other set of features) ensures a $\mathcal{CC}$ advantage remains to be addressed. 
 Given the significance of $\mathcal{OC}$ tasks, it might prove worthwhile to consider their generalizations to multipartite scenarios and explore potential application to the semi-device independent paradigm. 
 Another natural direction for future research is to look for information theoretic principles \cite{ic} that restrict success in $\mathcal{OC}$ tasks to quantum maximum.

\section{Acknowledgments} We thank M. Paw\l owski, M. Horodecki,  M. Oszmaniec and C. M. Scandolo for helpful discussion. 
This research was conducted in \textit{National Quantum Information Centre Gdansk}. 
This work is supported by NCN grants 2016/23/N/ST2/02817, 2014/14/E/ST2/00020 and FNP grant First
TEAM (Grant No. First TEAM/2016-1/5).

\newpage{\pagestyle{empty}\cleardoublepage}

\clearpage
\appendix
\onecolumngrid
\newpage
\section{Proofs of Lemmas}
In this appendix, we provide the proofs of all the \textit{Lemmas} used in the article.

\begin{lemma} \label{Ale:UpperBound}
For an instance of the subclass of $\mathcal{OC}$ tasks defined in section \ref{sec:oc}, the classical success probability $p_\mathcal{NC}$ is upper bounded in the following way,
\begin{align} \label{Ancb1}
p_{\mathcal{NC}} & \leq \max_{\{q_{a_1,a_2}\}} \bigg\{ \sum_b p(b) \max_c \bigg\{ 
\sum_{a_1,a_2} p(a_1|b) q_{a_1,a_2} \delta_{c,g(a,b)} \bigg\} \bigg\}
\end{align} 
where the outer maximization is over a set of variables $\{q_{a_1,a_2}\}$  satisfying the conditions,
\be \label{Aq1}
 q_{a_1,a_2} \geq0, \ \sum_{a_2} q_{a_1,a_2} = 1. \ee 
\end{lemma}
\begin{proof}
We follow the method introduced in \cite{shp}. Let us recall that the expression for maximal classical success probability for the $\mathcal{OC}$ task described in Eq. \eqref{oc} is,
\begin{align} \label{Aa3}
p_{\mathcal{NC}} = & \max_{\{\mathcal{E}\}\{\mathcal{D}\}} \bigg\{ \sum_{m} \sum_{b} p(b)  \bigg(\sum_{a} p(a|b) p_{\mathcal{E}}(m|a) p_{\mathcal{D}}(c=g(a,b)|b,m) \bigg)\bigg\} , 
\end{align}
where the message $m$ can take arbitrary number of distinct values. And we seek to obtain an upper bound of $p_{\mathcal{NC}}$ under the oblivious constraints,
\beq \label{Ac} 
\forall m,\forall a_1,a'_1 \in [n_{a_1}], \ p_{\mathcal{E}}(m) = p_{\mathcal{E}}(m|a_1) = \sum_{a_2} p(a_2|a_1) p_{\mathcal{E}}(m|a_1,a_2) = p_{\mathcal{E}}(m|a'_1) .
\eeq
We proceed in two steps: first, we observe that given an encoding scheme, the optimal decoding scheme $\mathcal{D}^*$ for $\mathcal{OC}$ task is fixed and deterministic. Then we provide a technique for recovering an upper bound on $p_{\mathcal{NC}}$ by finding the optimal encoding scheme $\mathcal{E}^*$ for a single level of the message.

\textit{Decoding in an $\mathcal{OC}$ task:} In-order to attain the maximal success probability, Bob's decoding strategy $p_{\mathcal{D}}(c|b,m)$ is to output the most probable value $g(a,b)$ given Alice's message $m$ pertaining to an encoding $\mathcal{E}$ and his input $b$. The right-hand-side of \eqref{Aa3} can be interpreted as the convex combination of elements $( \sum_{a} p(a,b) p_{\mathcal{E}}(m|a))$ with the weightage $p_{\mathcal{D}}(c|b,m)$ for each pair of $b,m$.
This in-turn implies that for a fixed encoding strategy Bob's optimal decoding strategy $\mathcal{D}^*$ is deterministic i.e.,
\beq \label{Ads}
&&p_{\mathcal{D}^*}(c|b,m) = 
\begin{cases}
1,  \text{if} \ \sum\limits_{a| g(a,b)=c} p(a,b) p_{\mathcal{E}}(m|a) \geq  \sum\limits_{a| g(a,b)\neq c} p(a,b) p_{\mathcal{E}}(m|a),\\
0,  \text{else}.
\end{cases}
\eeq
This allows us to re-express \eqref{Aa3} as,
\be \label{ANCB}
p_{\mathcal{NC}} = \max_{\mathcal{E}}\bigg\{ \sum_m \sum_{b} p(b) \max_{c} \bigg(\sum_{a} p(a|b) p_{\mathcal{E}}(m|a) \delta_{c,g(a,b)} \bigg) \bigg\}.
\ee 

\textit{Encoding in an $\mathcal{OC}$ task:} For any classical encoding strategy $\mathcal{E}$ define a set of non-negative parameters $\{q_{\mathcal{E},m}(a_1,a_2) := \frac{p(a_2|a_1) p_{\mathcal{E}}(m|a_1,a_2)}{p_{\mathcal{E}}(m)}\}$. It follows from the oblivious constraint \eqref{Ac} that,
\be \label{Aq}
\forall m,a_1, \ \sum_{a_2}  q_{\mathcal{E},m}(a_1,a_2) =1. 
\ee
 Using the additional condition $p(a_2|a_1,b)=p(a_2|a_1)$ we may now re-write (\ref{ANCB}) in terms of $q_{\mathcal{E},m}(a_1,a_2)$ as, 
\begin{align} 
p_{\mathcal{NC}} & 
=\max_{\mathcal{E}} \bigg\{ \sum_m p_{\mathcal{E}}(m) \sum_b p(b) \max_c  \bigg\{ \sum_{a_1,a_2} p(a_1|b) q_{\mathcal{E},m}(a_1,a_2) \delta_{c,g(a,b)} \bigg\} \bigg\}  \nonumber \\
& \leq \max_{\{q_{a_1,a_2}\}} \bigg\{ \sum_b p(b) \max_c \bigg\{ \label{AoptPc2} 
\sum_{a_1,a_2} p(a_1|b) q_{a_1,a_2} \delta_{c,g(a,b)} \bigg\} \bigg\}.
\end{align}
The last inequality is implied by the fact that $\sum_m p_{\mathcal{E}}(m)=1$. Specifically, the last inequality states that in-order to obtain an upper bound on $p_{\mathcal{NC}}$ its enough to find the optimal encoding strategy $\mathcal{E}^*$ for a single level of the message, which justifies the use of the symbol $q_{a_1,a_2}$. The constraint \eqref{Aq} along with the fact that $\forall a_1,a_2, \ q_{a_1,a_2} \geq 0$ implies that the set of all valid instances of $q_{a_1,a_2}$ form a convex polytope. Since the `max' function is convex, hence with regard to find a upper bound on  $p_\mathcal{NC}$ it is sufficient to evaluate the expression \eqref{AoptPc2} at the extremal points of that polytope and find the optimal.
\end{proof}

\begin{lemma} \label{Ale:extremal} The set of valid assignments of $\{q_{a_1,a_2}\}$ satisfying the linear constraints \eqref{Aq1} form a convex polytope. The extremal points of this polytope resemble deterministic probability distributions, i.e., any extremal point $\{q^{ext}_{a_1,a_2}\}$ is of the following form: for each $a_1$, $q^{ext}_{a_1,a_2}=0$ for all values of $a_2$ except a specific $\tilde{a_2}$ for which $q^{ext}_{a_1,\tilde{a_2}}=1$.
\end{lemma}
\begin{proof}
Let us represent the variables by a $n_{a_1} \times n_{a_2}$ matrix whose $(a_1,a_2)$-th element is $q_{a_1,a_2}$. Since $\sum_{a_2} q_{a_1,a_2} = 1$, each row of such matrix sums to 1.  The extremal points are described as follows. We consider a string  $(e_0,e_1,...,e_{n_{a_1}-1})$ where $e_{a_1} \in \{0,...,n_{a_2}-1\}$. Each extremal matrix is defined by this string such that $q_{a_1,a_2} = \delta_{a_2,e_{a_1}}$. There are $n_{a_2}^{n_{a_1}}$ number of such strings and each corresponds to an extremal point. One can check  that, any arbitrary matrix whose elements are $\tilde{q}_{a_1,a_2}$ can be obtained by the convex combination of these extremal points, in which the coefficient of the matrix corresponds to the string $(e_0,e_1,...,e_{n_{a_1}-1})$ is  $\prod^{n_{a_1}-1}_{i=0}\tilde{q}_{i,e_i}$. 
\end{proof}  


\begin{lemma} \label{ALemma3} For a given quantum prepare and measure communication complexity protocol the following holds,
\be 
\chi \leq d p_G,
\ee
where  $\chi = \sum_{x,y} p(x,y)\tr\left( M^y_{z=f(x,y)}\right)$, $d$ is dimension of the communicated system and $p_G$ is guessing probability without communication.
\end{lemma}
\begin{proof} It is straightforward to see that, when there is no communication, given $y$ the best strategy for Bob would be to output $f(x,y)$ which is more likely according to the prior probability of the inputs, i.e.,
\be \nonumber
p_G = \sum_y p(y) \max \left( \sum_{x|f(x,y)=0} p(x|y), \sum_{x|f(x,y)=1} p(x|y) \right)
\ee By denoting $\chi^y_z = \tr(M^y_z)$, and imposing the fact $\chi^y_0 + \chi^y_1 = d$, one obtains,
\beq 
& \chi & = \sum_{x,y} p(x,y) \chi^y_{z=f(x,y)} \nonumber \\
&& = d \sum_y p(y) \left( \sum_{x|f(x,y)=0} p(x|y) \frac{\chi^y_0}{d} +  \sum_{x|f(x,y)=1} p(x|y) \frac{\chi^y_1}{d} \right) \nonumber \\
&& \leq d \sum_y p(y) \max \left( \sum_{x|f(x,y)=0} p(x|y),  \sum_{x|f(x,y)=1} p(x|y)  \right) \nonumber \\
&& = d p_G .\nonumber
\eeq 
\end{proof}

\begin{lemma} \label{ALemma4} Given a $\mathcal{CC}$ problem and a protocol using a two-leveled classical message with a success probability $p_{\mathcal{C}_2}$, the success probability of a protocol using a $d$-leveled classical message is lower bounded in the following way,
\be \label{Apump}
p_{\mathcal{C}_d} \geq 1 - \exp{\left(-\frac{1}{2p_{\mathcal{C}_2}}\log{d}(p_{\mathcal{C}_2}-\frac{1}{2})^2\right)}.
\ee 
\end{lemma}
\begin{proof} We have a communication complexity protocol $\mathcal{P}$ which uses a bit of communication to obtain a success probability of $p_{\mathcal{C}_{2}}$. Now we shall use the \textit{pumping argument} to discern the desired thesis \eqref{Apump}. Consider yet another protocol $\mathcal{P}'$ wherein Alice and Bob repeat protocol $\mathcal{P}$ $\log{d}$ times. They produce as their final outcome the majority of outcomes obtained in $\log{d}$ runs of $\mathcal{P}$. If $\lceil \log{d} \rceil$ is even they succeed if $\mathcal{P}$ succeeds $\lceil \frac{\log{d}}{2} \rceil +1$ times and if $\lceil \log{d} \rceil$ is odd they succeed if $\mathcal{P}$ succeeds $\lceil \frac{\log{d}}{2} \rceil$ times. Consider the event that the protocol $\mathcal{P}$ succeeds and the number of simultaneous occurrence of such event is captured in the variable $\tau$. This allows us to lower bound $p_{\mathcal{C}_d}$ as,
\begin{align}
p_{\mathcal{C}_d}& \geq p\big(\tau>\left\lceil \frac{\log{d}}{2} \right\rceil\big) =\sum_{i=\lceil \frac{\log{d}}{2} \rceil +1}^{\lceil \log{d}\rceil} {{\lceil \log{d}\rceil}\choose{i}} p_{\mathcal{C}_2}^i(1-p_{\mathcal{C}_2})^{{\lceil \log{d}\rceil}-i} .\nonumber
\end{align}
The right hand side of the above equation is further lower bounded  based on Chernoff's inequality as,
\be \nonumber
p\big(\tau>\left\lceil \frac{\log{d}}{2}\right\rceil \big)\geq 1 - \exp{\left(-\frac{1}{2p_{\mathcal{C}_2}}\log{d}(p_{\mathcal{C}_2}-\frac{1}{2})^2\right)}. 
\ee
\end{proof}
\begin{lemma} \label{ALemma5} Given a $\mathcal{CC}$ problem and a protocol which achieves a success probability $p_{S}$ using $\mathcal{C}(f,p_\mathcal{S})$ bits, the success probability of a protocol using a two-leveled classical message is upper bounded in the following way,
\be \label{Apump2}
p_{\mathcal{C}_2}\leq \frac{1}{2}  + \sqrt{\frac{2p_S}{\mathcal{C}(f,p_S)}}.
\ee
\end{lemma}
\begin{proof} We have a communication complexity protocol which achieves success probability $p_S$ using $\mathcal{C}(f,p_S)$ bits of communication. We know from the \textit{pumping argument} used in the proof for \textit{Lemma 4},
\begin{align} 
p_{S} \geq 1 - \exp{\left(-\frac{1}{2p_{\mathcal{C}_2}}\mathcal{C}(f,p_S)(p_{\mathcal{C}_2}-\frac{1}{2})^2\right)}. \nonumber
\end{align}
Now expanding the above exponential term in the above inequality and taking the first two terms one retrieves,
\begin{align} 
p_{S} \geq {\left(\frac{1}{2p_{\mathcal{C}_2}}\mathcal{C}(f,p_S)(p_{\mathcal{C}_2}-\frac{1}{2})^2\right)}. \nonumber
\end{align}
This is conveniently re-expressed as,
\begin{align}
\frac{2p_S}{C(f,p_S)} &
\geq
\frac{(p_{\mathcal{C}_2}-\frac{1}{2})^2}{p_{\mathcal{C}_2}} \geq
(p_{\mathcal{C}_2}-\frac{1}{2})^2, \nonumber
\end{align}
where the second inequality follows 
from the observation that $0 \leq p_{\mathcal{C}_2}\leq 1$ and subsequently yields the desired thesis \eqref{Apump2}. 
\end{proof}
\begin{lemma} \label{ALemma6} For Hidden matching problem an $\mathcal{OC}$ task can be constructed with a success probability $p_{\mathcal{NC}}$, such that $p_\mathcal{NC} \leq p_{\mathcal{C}_2}$.
\end{lemma}
\begin{proof} In the hidden matching task, Alice is given a bit string $x \in \{0,1\}^n$ of length $n$ and Bob is given $y \in M_n$ where $M_n$ denotes the family of all
possible perfect matchings on $n$ nodes. Their goal is to output a tuple $z = ( i, j, t )$ such
that the edge $(i, j)$ belongs to the matching $y$ and $t = x_i \oplus x_j$. Being a relational problem, given an input $(x,y)$, Bob's task is to return $z$ from a set of possible relation, i.e., $R(x,y)= \{(i,j,t)\}$ such that $(i,j)\in y$ and $t=x_i\oplus x_j$. Subsequently, the success probability is given by $\sum_{x,y} p(x,y) \sum_{z\in R(x,y)} p(z|x,y)$, and in classical communication with two-dimensional system 
\beq \label{ACC2}
&p_{\mathcal{C}_2}&  = \max_{\{E\}\{D\}} \sum^1_{m=0} \sum_{y} p(y)  \bigg(\sum_{x,z\in R(x,y)} p(x|y) p_{E}(m|x) p_{D}(z|y,m)  \bigg) . 
\eeq
We follow the same construction of the $\mathcal{OC}$ task described in Fig. \ref{fig1}. The corresponding $\mathcal{OC}$ is also a relational problem in which
\be \nonumber
g(a,b) =
 \begin{cases}
 R(x,y) \quad \text{for } a_2=0, \\
\tilde{R}(x,y) \quad \text{for } a_2=1, 
\end{cases}
\ee where $ \tilde{R}(x,y)= \{(i,j,1\oplus t)\}$ such that $(i,j)\in y$ and $t= x_i\oplus x_j$. In other words, the hidden matching task is unaltered in the case of $a_2=0$, while for $a_2=1$, Bob's objective is to output one edge $(i,j)$ from the matching $y$ together with the complement of their xor, i.e., $i\oplus j \oplus 1$. Following \textit{Lemmas} \ref{Ale:UpperBound} and \ref{Ale:extremal} we first state the expression as given in \eqref{optnc*},
\beq  \nonumber
&p_{\mathcal{NC}} & \ = \max_{\{\mathcal{E}\}\{\mathcal{D}\}} \bigg\{ \sum_{m} \sum_{b} p(b)  \bigg(\sum_{a,c\in g(a,b)} p(a|b) p_{\mathcal{E}}(m|a) p_{\mathcal{D}}(c|b,m) \bigg)\bigg\} \\
&& \leq  \sum_{b} p(b) \sum_{a_1,a_2} p(a_1|b)\ \delta_{e_{a_1},a_2}\ \Delta_{c_b,g(a,b)} \ ,
\eeq where $\Delta_{c_b,g(a,b)}=1$ if $c_b\in g(a,b)$, otherwise 0.
Recall that in the proposed $\mathcal{OC}$ task $a_1=x,b=y,c=(i,j,t)$. Subsequently,  by denoting $c_y = (i^*,j^*,t^*)_y$ we re-write the above expression of $p_{\mathcal{NC}}$, 
\beq \label{Aeqhm}
&p_{\mathcal{NC}} \leq & \sum_y p(y) \sum_{x|{e_x}=0,(i^*,j^*,t^*)_y\in R(x,y)} p(x|y) + \sum_y p(y) \sum_{x|{e_x}=1,(i^*,j^*,t^*)_y\in \tilde{R}(x,y)} p(x|y) .
\eeq
Further, consider the following classical strategy employing two-leveled message $m\in \{0,1\}$,
\begin{align}
p_{E}(m|x) = \delta_{m,e_x} \ , \ p_{D}(i,j,t|y,m) =
    \delta_{(i,j,t),(i^*,j^*,m \oplus t^*)_y}\ . \nonumber
\end{align}
Inserting this strategy in \eqref{ACC2}, and using the following feature of hidden matching problem,
\be \nonumber 
\forall \ y,(i,j,t), \quad \sum_{x|(i,j,1\oplus t)\in R(x,y)} p(x|y) = \sum_{x|(i,j,t)\in \tilde{R}(x,y)} p(x|y) ,\ee
one obtains the same expression of success probability in $\mathcal{CC}$ problem as given in the right side of \eqref{Aeqhm},
\beq
&p_{\mathcal{C}_2} \geq & \sum_y p(y) \sum_{x|m=0,(i^*,j^*,t^*)_y\in R(x,y)} p(x|y)  + \sum_y p(y) \sum_{x|m=1,(i^*,j^*,1\oplus t^*)_y\in R(x,y)} p(x|y)  \nonumber \\
&= & \sum_y p(y) \sum_{x|m=0,(i^*,j^*,t^*)_y\in R(x,y)} p(x|y)  + \sum_y p(y) \sum_{x|m=1,(i^*,j^*,t^*)_y\in \tilde{R}(x,y)} p(x|y)  \nonumber \\
& \geq & p_{\mathcal{NC}} . \nonumber
\eeq

Note that to show the quantum advantage in the $\mathcal{OC}$ task, we consider the same quantum strategy as described in \textit{Result} \ref{result:1} which leads to $p_{\mathcal{Q}} = \frac{1}{d}(2p_{\mathcal{Q}_d} + d - 1 - \chi)$ where $\chi = \sum_{x,y,z\in R(x,y)} p(x,y) \tr(M^y_z)$. Subsequently, one can show the validity of \textit{Corollary} \ref{coro3}. 

\end{proof}



\begin{thebibliography}{100}
\bibitem{comment} To avoid confusion, note that, `communication tasks' referred in \cite{chutiyapa} are solely based on oblivious constraints. In this letter, we call the same as `oblivious communication tasks'.
\end{thebibliography}
\end{document}